\input{aipcheck}

\documentclass[
    ,final            
  ]
  {aipproc}

\layoutstyle{6x9}

\begin{document}

\title{Neutrino probe of cosmic ray astrophysics  and new physics at sub-fermi distances}


\author{Luis A. Anchordoqui  [for the Pierre Auger Collaboration]}{
  address={Department of Physics, University of Wisconsin-Milwaukee,
Milwaukee, WI 53201}
}

\begin{abstract} We show that if the evolution of cosmic ray sources
  follows that of active galactic nuclei, the upper limit on the
  diffuse flux of tau neutrinos from the Pierre Auger Observatory
  marginally constrains the proton fraction at the end of the energy
  spectrum.  We also discuss prospects to uncover new physics
  leptophobic interactions using future Auger data.  \end{abstract}

\maketitle



The Pierre Auger Observatory is a major international effort aiming at
high-statistics study of highest energy cosmic rays to uncover their
origins and nature~\cite{Abraham:2004dt}. These cosmic rays are too
rare to be directly detected, but the direction, energy, and to some
extent the nature of the primary particle can be inferred from the
cascade of secondary particles induced when the primary impinges on
the upper atmosphere.  These cascades, or air showers, have been
studied in the past by measuring the nitrogen fluorescence they
produce in the atmosphere, or by directly sampling shower particles at
ground level. Auger is a ``hybrid'' detector, exploiting both of these
methods by employing an array of water \v{C}erenkov detectors
overlooked by fluorescence telescopes; on clear, dark nights air
showers are simultaneously observed by both types of detectors,
facilitating powerful reconstruction methods and control of the
systematic errors which have plagued cosmic ray experiments to date.
Auger also provides a promising way of detecting ultra-high energy
neutrinos by looking for deeply--developing, large zenith angle
($>60^\circ$) or horizontal air showers. At these
large angles, hadronic showers have traversed the equivalent of
several times the depth of the vertical atmosphere and their
electromagnetic component has extinguished far away from the
detector. Only very high energy core--produced muons survive past 2
equivalent vertical atmospheres. Therefore, the shape of a hadronic
(background in this case) shower front is very flat and very prompt in
time. In contrast, a neutrino shower appears pretty much as a
``normal'' shower. It is therefore straightforward to distinguish
neutrino induced events from background hadronic showers. Moreover, because of 
full flavor mixing, tau neutrinos could be as abundant as
other species and so very low $\nu_\tau$ fluxes could be detected very
efficiently by Auger detectors by looking at the interaction in the
Earth crust of quasi horizontal $\nu_\tau$ inducing a horizontal
cascade at the detector.

The Pierre Auger Observatory has confirmed a strong steepening of the
cosmic ray flux above
$10^{10.6}$~GeV~\cite{Abraham:2008ru,Abraham:2009wk}. If such a
steepening is due to cosmic ray interactions with the cosmic micorwave
background, the cosmic ray arrival directions at the energies above
the suppression are expected to correlate with the nearby matter
distribution, which is quite inhomogeneous. Observation of these
correlations would provide a doorway towards cosmic ray astronomy
(provided that particle deflections in the intervening galactic and
extragalactic magnetic fields are relatively small). Anisotropy clues
have been already reported in the data: cosmic ray events above
$\simeq 10^{10.8}$~GeV seem to correlate over angular scales of less
than $6^\circ$ with the directions of nearby (distance < 100 Mpc)
active galactic nuclei~\cite{Cronin:2007zz}, though the evidence for
such anisotropy has not strengthened in a more recent analysis with
larger statistics~\cite{Abraham:2009eh}. In addition, upper limits on
the diffuse neutrino flux and the ultra-high energy cosmic ray photon
fraction have been established~\cite{Abraham:2009eh}. All these
observations strongly support the hypothesis that cosmic ray
acceleration takes place in astrophysical objects.  The measurements
of the variation of the depth of shower maximum with energy,
interpreted with current hadronic interaction models, favour a mixed
cosmic ray (protons + nuclei) composition at energies above
$10^{8.6}$~GeV~\cite{Abraham:2009ds}.  It is interesting to note,
however, that uncertainties in the proton-air cross-section at the
high energies characteristic of air showers may be large enough to
alter this conclusion~\cite{Ulrich:2009yq}. In what follows we show
that limits on neutrino fluxes provide complementary information to
ascertain the cosmic ray nature.

It is helpful to envision the cosmic ray engines as machines where
protons are accelerated and (possibly) permanently confined by the
magnetic fields of the acceleration region. The production of neutrons
and pions and subsequent decay produces neutrinos, gamma rays, and
cosmic rays. If the neutrino-emitting source also produces high energy
cosmic rays, then pion production must be the principal agent for the
high energy cutoff on the proton spectrum.  Conversely, since the
protons must undergo sufficient acceleration, inelastic pion
production needs to be small below the cutoff energy; consequently,
the plasma must be optically thin. Since the interaction time for
protons is greatly increased over that of neutrons because of magnetic
confinement, the neutrons escape before interacting, and on decay give
rise to the observed cosmic ray flux. The foregoing can be summarized
as four conditions on the characteristic nucleon interaction time
scale $\tau_{\rm int}$; the neutron (pion, muon) decay lifetime
$\tau_{n(\pi,\mu)}$; the characteristic cycle time of confinement
$\tau_{\rm cycle}$; the cooling time scale of charged particles
$\tau_{\rm cool}$; and the total proton confinement time $\tau_{\rm
  conf}$: $(i)\; \tau_{\rm int}\gg \tau_{\rm cycle}$; $(ii)\; \tau_n >
\tau_{\rm cycle}$; $(iii)\; \tau_{\rm int}\ll \tau_{\rm conf}$;
$(iv)\; \tau_{\pi,\mu} < \tau_{\rm cool}$. The first condition ensures
that the protons attain sufficient energy.  Conditions $(i)$ and
$(ii)$ allow the neutrons to escape the source before
decaying. Condition $(iii)$ permits sufficient interaction to produce
neutrons and neutrinos. Condition ($iv$) ensures that cooling
processes of secondary pions, and muons have negligible influence on
the energy distribution of the emitted neutrons and neutrinos.

Within this working hypothesis the relative number and energy of the
neutrinos and neutrons depend only on kinematics, which implies
approximate equipartition of the decaying pion's energy between the
neutrinos and the electron, and the relative radiation density in the
source. On average, each interaction will produce $\eta$
neutrinos per neutron with relative energy $\epsilon$ per neutrino,
{\it i.e.}
\begin{equation}
  \eta = \frac{\langle N_\nu\rangle}{\langle N_n\rangle}\quad 
{\rm and}\quad\epsilon = 
  \frac{\langle E_\nu\rangle}{\langle E_n\rangle}\,.
\end{equation}
The neutrino emissivity of flavor $i$ is then given by:
\begin{equation}
\frac{\Delta E_{\nu_i}}{N_{\nu_i}}\mathcal{L}_{\nu_i}(z,E_{\nu_i})
= \frac{\Delta E_n}{N_n}\mathcal{L}_n(z,E_n)\,.
\end{equation}
Assuming flavor universality as well as $\epsilon \simeq E_\nu/E_n
\simeq \Delta E_\nu/\Delta E_n$ and $\eta \simeq
N_{\mathrm{all}\,\nu}/N_n$ we arrive at the neutrino source luminosity
(per co-moving volume):
\begin{equation}\label{ratio}
\mathcal{L}_{\mathrm{all}\,\nu}(z,E_\nu) \simeq\frac{\eta}{\epsilon}\,
\mathcal{L}_n(z,E_\nu/\epsilon)\,.
\end{equation}
In the following, we will consider a (hypothetical) source, where pion
production proceeds exclusively via resonant $p\gamma \to \Delta^+$
scattering with fixed values $\eta=3$ and $\epsilon=0.07.$ Note, that
the relation~(\ref{ratio}) derived for optically thin sources can be
regarded as a {\em lower} limit on the neutrino luminosity as long as
energy-loss processes in the source are negligible.  Next, we
translate this conservative expectation into an {\em upper} limit on
the extra-galactic proton fraction in ultra-high energy cosmic rays,
exploiting the experimental upper limit on the diffuse flux of tau
neutrinos~\cite{Abraham:2009eh}.

\begin{figure}
  \includegraphics[height=.3\textheight]{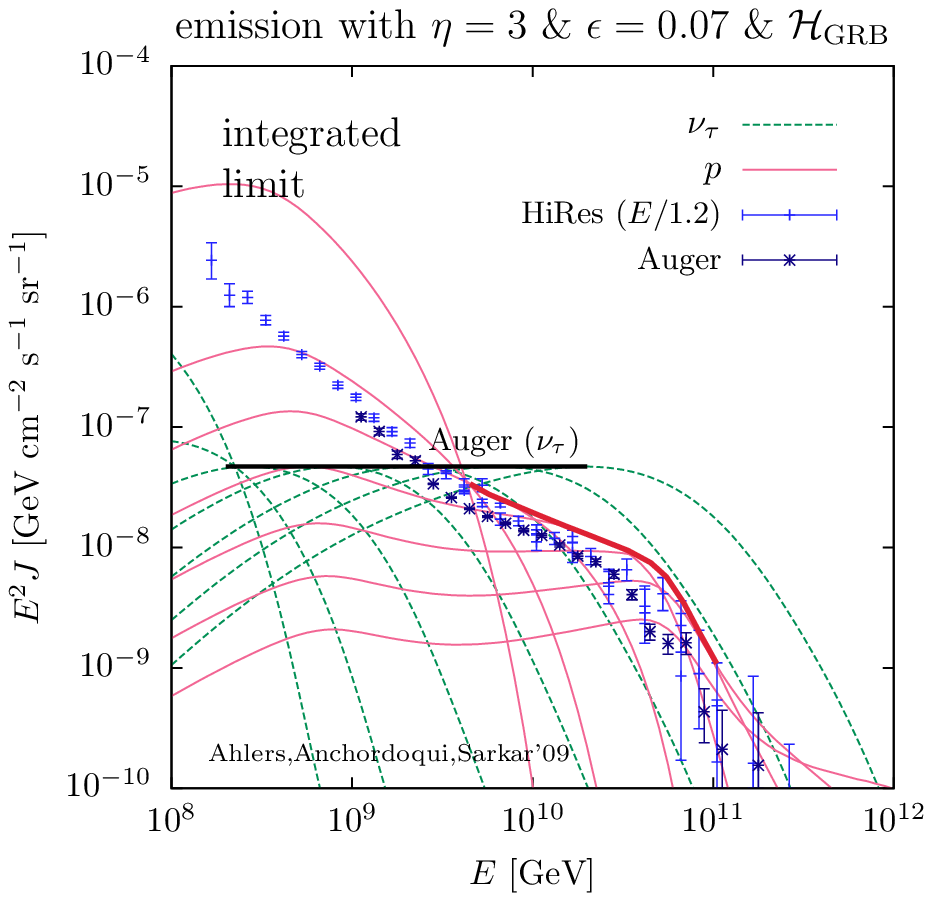}
  \includegraphics[height=.3\textheight]{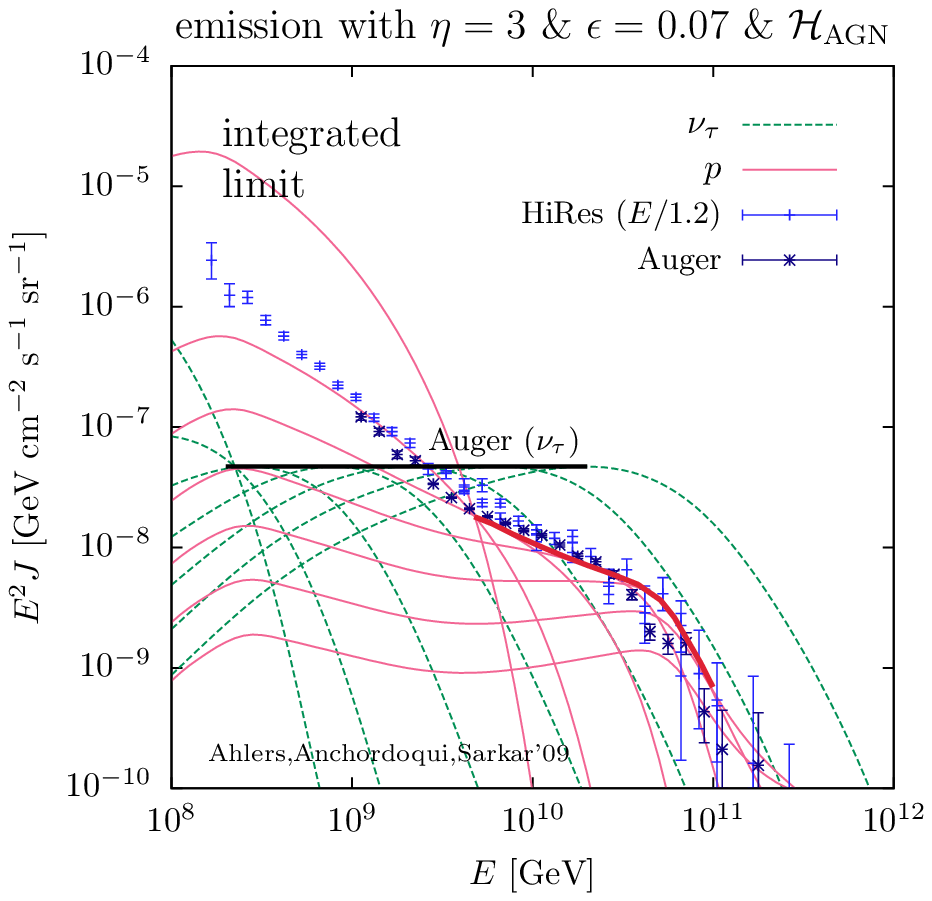}
  \caption{Upper limits on the proton contribution in ultra-high
    energy cosmic rays derived from Auger limit on the diffuse $\nu_\tau$-flux 
 in models where the cosmic ray sources evolve
    strongly with redshift; we show the case of strong cosmological
    evolution of the proton sources according to $\mathcal{H}_{\rm
      GRB}$ and $\mathcal{H}_{\rm AGN}$.}
\label{diagnostics}
\end{figure}

For this procedure we introduce test functions of the neutron source
luminosity of the form $\mathcal{L}^{\rm test}_n(0,E) = \mathcal{L}_0
\, (E/E_\mathrm{max})^{-1}\, \exp (-E/E_\mathrm{max})\,,$ with an
exponential energy cut-off $E_\mathrm{max}$ that we vary between
$10^{8}$~GeV and $10^{12}$~GeV with a logarithmic step-size of
$\log_{10}E = 0.25$. We adopt the usual concordance cosmology of a
flat universe dominated by a cosmological constant with
$\Omega_{\Lambda} \sim 0.7$, the rest being cold dark matter with
$\Omega_\mathrm{m} \sim 0.3$. The Hubble parameter is given by
\mbox{$H^2 (z) = H^2_0\,(\Omega_{\mathrm{m} } (1 + z)^3 +
  \Omega_{\Lambda})$}, normalized to its value today of 70
km\,s$^{-1}$\,Mpc$^{-1}$. The time-dependence of the red-shift can be
expressed via $\mathrm{d}z = -\mathrm{d} t\,(1+z)H$. The cosmological
evolution of the source density per co-moving volume is parameterized
as $ \mathcal{L}_i(z,E) = \mathcal{H}(z)\mathcal{L}_i(0,E)\,,$ where
the source luminosity per co-moving volume is assumed to follow that
of gamma ray bursts $\left[ \mathcal{H}_{\rm GRB}(z) = (1+z)^{4.8}
\right.$, for $z<1$; $\mathcal{H}_{\rm GRB}(z) = N_1 (1 +z)^{3.1}$, for
$1<z<4$;  $\mathcal{H}_{\rm GRB}(z) = N_1 N_4 (1+z)^{-0.1}$, for
$z>4$; with $N_1 = 2^{3.7}$ and $\left. N_4 = 5^{3.2} \right]$ and
that of active galactic nuclei $\left[\mathcal{H}_{\rm AGN}(z) =
  (1+z)^{5.0} \right.$, for $z<1.7\,$; $\mathcal{H}_{\rm AGN}(z) =
N_{1.7}$, for $1.7<z<2.7\,$;  $\mathcal{H}_{\rm AGN}(z) =
N_{1.7}\,N_{2.7}^{(2.7-z)}$, for $z>2.7$; with $N_{1.7}=2.7^5$ and
$\left. N_{2.7}=10^{0.43}\right]$. Each neutron test luminosity is
related to a neutrino luminosity by the ratio~(\ref{ratio}). After
propagation we normalize the accumulated contribution of
extra-galactic and cosmogenic neutrinos to the limit on ultra-high
energy tau neutrinos from Auger. For details on the calculation
see~\cite{Ahlers:2009rf}. The results are shown in
Fig.~\ref{diagnostics}. For $\mathcal{H}_{\rm AGN}(z)$, the upper
limit on the diffuse flux of tau neutrinos marginally constrains the
proton fraction somewhat around $10^{10}~{\rm GeV}.$ 


Cosmic neutrinos are also unique probes of new physics as their
interactions are uncluttered by the strong and electromagnetic forces
and can reach center-of-mass energies $\sqrt{s} \sim
245$~TeV. However, rates for new physics processes are difficult to
test since the flux of cosmic neutrinos is virtually unknown.  It is
possible to mitigate this by using multiple observables which allow
one to decouple effects of the flux and cross section. In particular,
possible deviations of the neutrino nucleon cross section due to new
physics leptophobic interactions (e.g., production and decay of
TeV-scale black holes) can easily be uncovered at Auger by combining
information of the two different neutrino
channels~\cite{Anchordoqui:2001cg}. Namely, if an anomalously large
quasi-horizontal deep shower rate is found, it may be ascribed to
either an enhancement of the incoming neutrino flux, or an enhancement
in the neutrino-nucleon cross section.  However, these two
possibilities may be distinguished by separately binning events which
arrive at very small angles to the horizontal, the so-called tau
Earth-skimming events. An enhanced flux will increase both
quasi-horizontal and Earth-skimming event rates, whereas a large black
hole cross section suppresses the latter, because the hadronic decay
products cannot escape the Earth crust. This argument requires only
counting experiments and does not rely on measurements of shower
properties.

In summary, statistics and better experimental handles should enable
us to reconstruct the high end of the energy spectrum, to locate the
cosmic ray sources in the sky, and to discern the primary mass
composition. Future Auger data will not only provide clues to the
cosmic ray origin, but could enhance our understanding of fundamental
physics at sub-fermi distances. An optimist might even imagine the
discovery of microscopic black holes, the telltale signature of the
universe's unseen dimensions.


\begin{theacknowledgments}
  I would like to thank Markus Ahlers, Jonathan Feng, Haim Goldberg,
  Subir Sarkar, and Al Shapere for the most enjoyable
  collaborations. This work has been partially supported by NSF Grant
  No PHY-0757598 and the UWM Research Growth Initiative.
\end{theacknowledgments}



\bibliographystyle{aipproc}   

\bibliography{sample}

\IfFileExists{\jobname.bbl}{}
 {\typeout{}
  \typeout{******************************************}
  \typeout{** Please run "bibtex \jobname" to optain}
  \typeout{** the bibliography and then re-run LaTeX}
  \typeout{** twice to fix the references!}
  \typeout{******************************************}
  \typeout{}
 }


\end{document}